\documentclass{emulateapj}
\usepackage{graphicx}
\def\beq{\begin{eqnarray}}
\def\eeq{\end{eqnarray}}
\def\nserc{n_{s}}
\def\gamlam{\gamma_{\lambda}}
\def\neff{n_{\rm eff}}

\begin{document}

\title{The Intrinsic Properties of SDSS Galaxies}

\author{Ariyeh H. Maller}
\affil{Department of Physics, New York City College of Technology, CUNY, 300 Jay St., Brooklyn, NY 11201}
\author{Andreas A. Berlind}
\affil{Department of Physics and Astronomy, Vanderbilt University, 1807 Station B, Nashville, TN  37235}
\and
\author{Michael R. Blanton, David W. Hogg}
\affil{Center for Cosmology and Particle Physics, Department of Physics, 
New York University, 4 Washington Place, New York, NY 10003}
\begin{abstract}
The observed properties of galaxies vary with inclination; for most applications 
we would rather have properties that are independent of inclination, 
\emph{intrinsic} properties.  One way to determine inclination corrections is to 
consider a large sample of galaxies, study how the \emph{observed} properties 
of these galaxies depend on inclination and then remove this dependence to 
recover the \emph{intrinsic} properties.  We perform such an analysis for 
galaxies selected from the Sloan Digital Sky Survey which have been matched
to galaxies from the Two-Micron All Sky Survey.  We determine inclination corrections 
for these galaxies as a function of galaxy luminosity and Sersic index.  In the 
$g$-band these corrections reach as as high as 1.2 mag and have a median value of 
$0.3$ mag for all galaxies in our sample.
We find that the corrections show little dependence on galaxy luminosity,
except in the $u$ band, but are strongly dependent on galaxy Sersic index.

We find that the ratio of red-to-blue galaxies changes from 1:1 to 1:2 
when going from observed to intrinsic colors for galaxies in the range 
$-22.75 < M_K < -17.75$.  We also discuss how survey 
completeness and photometric redshifts should be determined when taking into 
account that observed and intrinsic properties differ.  Finally, we examine 
whether previous determinations of stellar mass give an intrinsic quantity or 
one that depends on galaxy inclination.
\end{abstract}


\section{Introduction}
In our search to understand the formation and evolution of galaxies, 
some of our primary tools
are measurements of the distribution of galaxy properties and relationships among these 
properties.  From observations of these quantities at different redshifts we can deduce 
the nature of galactic evolution and by comparing them to the properties of dark matter 
halos we can constrain models of galaxy formation.  

The measurement of galactic distributions includes, but is not limited to: the galaxy 
luminosity function \citep{hubb:36}, the galaxy correlation function 
\citep[the distribution of galaxies' spatial separations]{ph:74}, 
the galaxy velocity function \citep{gonz:00}, and the distributions of galaxy
sizes \citep{chol:85}, surface brightnesses \citep{free:70}, colors \citep{baum:59,faber:73}, 
metalicities \citep{oste:70} and star formation rates \citep{td:80}.  
Relationships between galaxy properties include, the Tully-Fisher relation \citep{tf:77}, 
the Faber-Jackson \citep{fj:76} and fundamental plane \citep{dd:87,dres:87} relations, 
the luminosity-size relation \citep{korm:77}, the luminosity-metalicity relation
\citep{faber:73, lequ:79} and the density-morphology relation \citep{dres:80}.

However, with the notable exception of the Tully-Fisher relation these distributions and
relations are traditionally measured in terms of the {\it observed} properties of galaxies.
That is, the 
measurements used are K-corrected and corrected for foreground dust extinction, but no 
correction is attempted to compensate for the viewing angle from which the galaxies are 
observed.  In contrast, the Tully-Fisher relation is not a relationship between a 
galaxy's \emph{observed} luminosity and rotation velocity, but a relation between a 
galaxy luminosity and rotation velocity \emph{corrected for inclination}.  The 
inclination correction attempts to recover the intrinsic properties of a galaxy and 
not properties that are measured because of the particular angle from which the galaxy is 
viewed. Spiral galaxies are observed to have redder colors when their disks are 
more inclined, which is expected if the inclination increases the amount of dust that 
light traverses when emitted from the galaxy.

Clearly, we would prefer to measure all galactic distributions and relationships in terms 
of intrinsic galaxy properties instead of observed ones.  The comparison of theory to 
observations is complicated and often done incorrectly because of confusion between 
observed and intrinsic galaxy properties.  Early semi-analytic models were unable to 
match both the galaxy luminosity function and the Tully-Fisher relation in part because 
they failed to take into account that the first is observed luminosity while the second is 
intrinsic luminosity \citep{sp:99}.  Also, when comparing galaxies at different 
redshifts we would like to be able to distinguish between evolution in their stellar 
populations and changes in their dust properties.  

Furthermore, inclination effects are of great help in understanding 
the nature of dust in galaxies.  Theoretical modeling of attenuation in galaxies is 
complicated because it not only depends on the properties of dust, which seem to vary 
between galaxies, but also on how the dust is distributed and mixed with stars.  By 
determining the intrinsic properties of galaxies we also learn how those properties 
change as a function of galaxy inclination and therefore some properties of the dust
distribution.   

To determine intrinsic properties we need to know how a galaxy's properties change as 
a function of its inclination.  This is different then removing the effects of dust and dust is
still present for a face-on galaxy.  There are a number of approaches for addressing this 
issue each of which has its own merits and disadvantages.  One approach is to solve for 
an inclination correction that minimizes the scatter in the Tully-Fisher relation 
\citep[e.g.,][]{verh:01}, which assumes that the scatter in this relation should be 
as small as possible.  Another method is to fit stellar population models to the SED 
of a galaxy and then assume that any discrepancies are 
caused by dust \citep[e.g.,][]{kauf:03}. A third is to observe background objects 
behind a foreground galaxy to get a direct measure of the extinction through the 
galaxy \citep[e.g.,][]{berl:97,holw:05}, but this is difficult to do for more than 
a handful of cases.  Finally, one can simulate the radiative transfer through a 
galaxy \citep[e.g.,][]{rocha:07}, assuming one knows the distribution and scattering 
properties of the dust.

The approach we explore here is somewhat simpler in that it assumes no knowledge of 
stellar population or dust properties.  Instead, the main assumption is that a galaxy's 
properties should be independent of inclination.  Thus any statistical 
correlation between a galaxy property and inclination can be attributed to dust and 
the inclination correction is whatever makes the observed correlation go away.  
This procedure has been applied a number of times  
\citep{giov:94,giov:95,tully:98,mgh:03,shao:07}.  In this paper we greatly expand 
upon this method by applying it to 10,340 galaxies taken from the Sloan Digital Sky 
Survey \citep[SDSS,][]{york:00} with accompanying infrared magnitudes from the 
Two-Micron All Sky Survey \citep[2MASS,][]{skru:06}.  It is important to have galaxies 
with near infrared photometry because the effects of attenuation are minimized in these 
wavebands \citep{bd:01}.  In a subsequent paper we will extend the analysis performed 
here to the full SDSS galaxy catalog.

We describe the method for determining inclination corrections in \S\ref{sec:method}.  
In \S\ref{sec:data} we describe the sample we will use and discuss some of the 
properties of galaxies in this sample.  In \S\ref{sec:results} we determine 
inclination corrections using our sample and compare our results to other 
determinations in \S\ref{sec:compare}.  In \S\ref{sec:changes} we discuss how 
consideration of intrinsic properties can change our conclusions about the distribution
of galaxy properties focusing on the color-magnitude diagram.  We also comment on
the effect on survey completeness, stellar masses and photometric redshifts.
\S\ref{sec:conc} contains our conclusions and some discussion of future directions.

\section{The Method}
\label{sec:method}
The method of determining extinction corrections statistically is based on looking for 
correlations between a galaxy property and galaxy inclination.  Under the assumption 
that the intrinsic properties of galaxies do not depend on inclination, one can infer the
effect of attenuation by plotting how an observed galaxy property changes
with inclination. This is shown in Figure \ref{fig:inc} for the case of galaxy color,
the property we will focus on in this paper.  The extinction correction then is
whatever is needed to remove the observed correlation between the chosen property 
and galaxy inclination.  Note that this procedure only measures attenuation relative 
to face-on galaxies, it can not say anything about the total attenuation that occurs 
in a galaxy.
\footnote{It is also possible that an observed correlation between a galaxy
property and inclination is the result of the data reduction.  Nonetheless, in this
case we would still like to correct this bias.  The interpretation that the correlation 
is caused by dust therefore does require study by analyzing the data pipeline and 
comparing to expectations from dust modeling. For simplicity, in this paper we will assume that dust is the dominate source of any observed correlations.}

This method has been applied to late type galaxies where the property observed to vary 
with inclination is magnitude in an isophotal radius, galaxy color, or the galaxy luminosity 
function \citep{giov:94,giov:95,tully:98,mgh:03,shao:07}.
In order to describe these various studies within one framework we turn to a more 
mathematical description of the procedure.

\begin{figure} [t]
   \centering
   \includegraphics[width=0.48\textwidth]{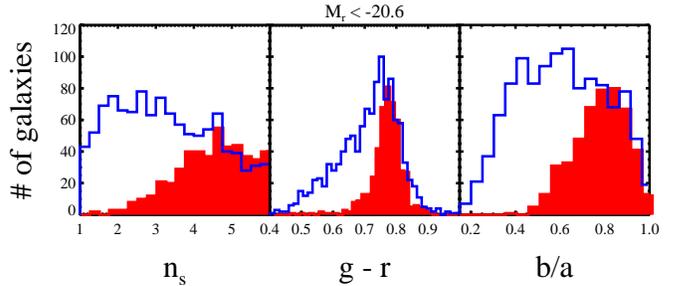} 
   \vspace{-1cm}
   \caption{Histograms of visually classified elliptical (shaded) and disk (line) galaxies 
   are shown in Sersic index, $\nserc$, $g - r$ color and axis ratio, $b/a$. In each case 
   the distribution of elliptical and disk galaxies are very different.  Often Sersic index
   or color is used to divide elliptical and disk galaxies, but as is clear from the figure 
   this still leaves many disks galaxies on the wrong side of the divide.  Axis ratio can be
   used as a strong discriminator of galaxy type and, when combined with Sersic index, 
   gives a sample that includes $70\%$ of all disk galaxies.  
   }\label{fig:type}
\end{figure}

Galaxy surveys measure a number of galaxy properties; fluxes, redshift, surface brightness, 
half light radius, axis ratio, position angle, etc.  Fluxes are usually converted 
into luminosities, a quantity that doesn't depend on the galaxy's distance 
(it has translational invariance) using the galaxy's redshift and a K-correction.
These constitute a set of observed galaxy data, 
$G^o = \{M^o_{\lambda}, \mu^o, r^o_{50},b/a, pa$, etc$\}$.  However, we would like to know 
the intrinsic properties of the galaxy, $G^i$, those properties that are invariant 
to rotation and translation.  We assume that the two are related by a transformation, 
$T$, such that
\beq
G^o = T(\theta,G^i) G^i
\eeq
where $T$ depends on the inclination angle of the galaxy, $\theta$, and possibly on other galaxy 
properties.  If intrinsic galaxy properties do not depend on inclination then we can 
determine the intrinsic galaxy properties by solving for the inverse transformation 
$T^{-1}$ as the operation that satisfies, 
\beq
\label{eq:main}
{{\partial (T^{-1}G^o)}\over{\partial \theta}}\Big |_{G^i} = 0
\eeq
Note that this derivative is with respect to intrinsic galaxy properties.
In practice this reduces to assuming a functional form for $T^{-1}$ and then optimizing
the parameters in that function to come as close as possible to equation \ref{eq:main}.

\begin{figure*} 
   \centering
   \includegraphics[width=\textwidth]{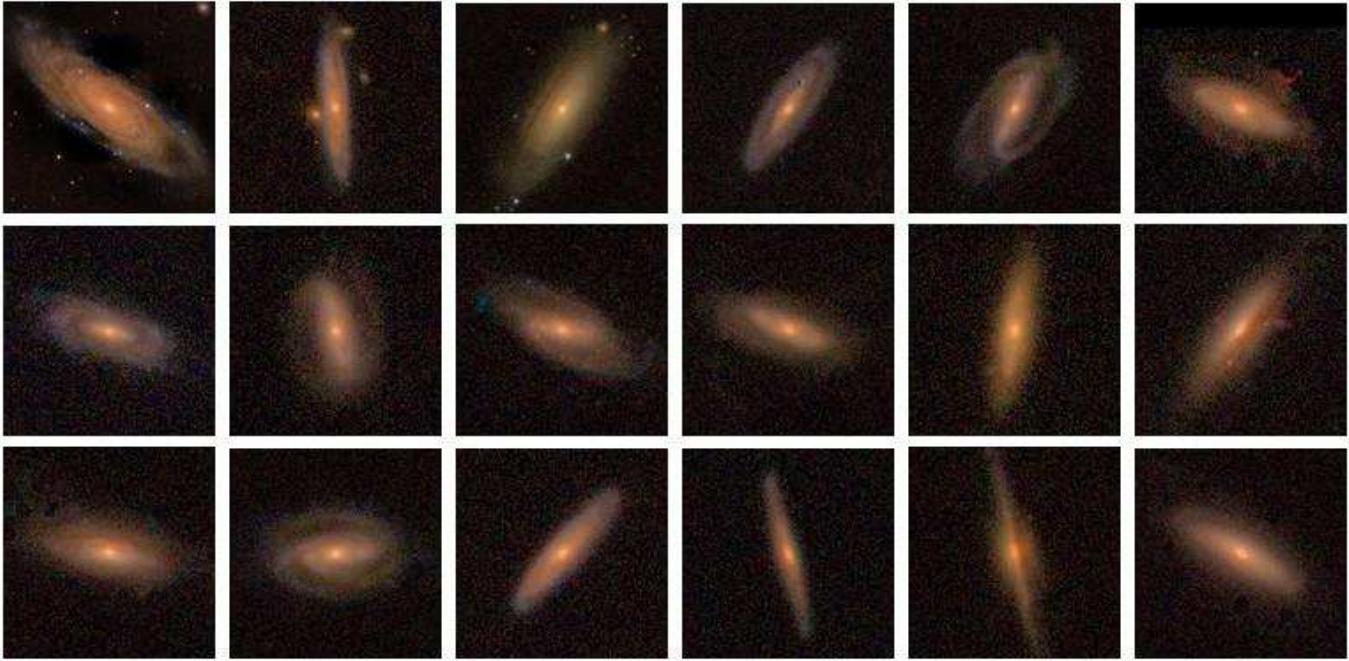}
   \vspace{0cm}
   \caption{Images of red ($g - r > 0.75$) and concentrated ($\nserc > 4.0$) and inclined 
   ($b/a < 0.5$)  galaxies from our sample.  Clearly these red, concentrated galaxies are 
   disk galaxies.  Besides the cuts stated the only selection in these images is that the 
   galaxy has a large angular size so that it makes a nice image.  Axis ratio is an 
   important diagnostic of galaxy type.
   }\label{fig:reddisks}
\end{figure*}

One's choice about the functional form of $T^{-1}$ therefore ends up having a strong 
effect on the conclusions reached.  Often it is assumed that the attenuation in a 
given waveband, $A_{\lambda}$, is of the form 
\beq
\label{eq:old}
A_{\lambda} = -\gamlam \log(\cos\theta)
\eeq
\citep[][and others]{giov:94}, where $\gamlam$ 
represents the combination of dust properties, the distribution of dust and the 
distribution of stars that determines the attenuation at a given inclination.
However, there is evidence that the dependence on inclination may be better fit 
by $\log^2(\cos\theta)$ \citep{mgh:03, rocha:07}.  The parameter gamma was taken to 
be a constant for a given wavelength in early work \citep{giov:94}, but subsequently 
has been considered as a function of luminosity \citep{giov:95,tully:98,mgh:03,shao:07}.
However, without a full understanding of the nature and distribution of dust in 
galaxies we can hardly know what galaxy properties $\gamlam$ should depend on.  
We must rely on the data and explore what functional forms and dependencies fit best.

\section{The Data}
\label{sec:data}
Our galaxy sample has been taken from the NYU-VAGC \citep{blan:03c}.  We use galaxies
from SDSS-DR2 that have near infrared magnitudes from 2MASS 
\citep[for a discussion of the nature and completeness of these galaxies see][] {mbwk:06}.  
Each galaxy has  been fit in the $r$-band with a Sersic profile using elliptical isophotes giving  us a Sersic index, $\nserc$, a half light radius $r_{50}$ and an axis-ratio $b/a$.
We also have $u,g,r,i$ and $z$ total magnitudes from SDSS and $J, H$ and  $K_s$ 
total magnitudes from 2MASS.  In order to insure that our inclination measurements are 
accurate we restrict our sample to galaxies with a seeing-deconvolved half light radius 
of three or more pixels. 
We also restrict ourselves to fits that are within the bounds of the allowed parameter 
space and not at one of the limits.  Thus we only include galaxies with Sersic index and
inclination in the range $0.5 < \nserc < 6.0$ and $0.15 < b/a < 1.0$.  Finally we restrict 
our sample to $-17.75 \ge M_K \le -22.75$ and $r_{50} \le 15$kpc, the range where we 
have enough galaxies to make statistical statements about our sample.  With these cuts 
we are left with $10,340$ galaxies.  Furthermore, $1,634$ of these galaxies with 
$M_r < -20.6$ have been visually classified by one of us (MRB).  Because of the $r$ selection 
of SDSS and K selection of 2MASS this is not a flux limited sample and thus we can not 
discuss the space densities of these galaxies.  Therefore we will not discuss space 
volume at all in this paper, but will defer such discussion to a following paper that 
expands on our treatment here to the full SDSS galaxy catalog.

\subsection{Disk Galaxies}
\label{ssec:disks}
As described in \S\ref{sec:method} the method is based on identifying observed properties 
that depend on inclination and then removing this dependency.  However, inclination is not a
directly observable galaxy property.  Instead, what one measures is axis ratio.  
Previous studies 
have attempted to convert axis ratio to inclination assuming a thickness for a galaxy's 
disk.  Before one can even do this though, one has to first identify which galaxies are disk
galaxies.  Usually this is done by cutting the sample by concentration, Sersic index or 
color.  Fig \ref{fig:type} shows histograms of the 1634 visually classified galaxies with
$M_r < -20.6$ in Sersic index, $g - r$ color and axis ratio separately for disk (solid line)
and elliptical (shaded region) galaxies.  We include S0 galaxies as disk galaxies because 
they have a disk and therefore their measured axis ratio should be more related to the disks 
inclination then to the ellipticity of the spheroid component.  It is clear that almost pure 
disk samples can be gotten by restricting one's sample to $\nserc < 3.0$ or $g - r < 0.7$; 
however, in both these cases close to half of the disk galaxies are not included in the 
sample.  Combining these two requirements leaves to a very mild improvement in sample 
completeness of only $\sim5\%$.
This is surprising since such cuts are often used in the literature to separate early and 
late type galaxy populations.  It is worth noting that when such a cut is employed roughly 
half of the galaxies designated as early type in our sample are in fact disk galaxies.
\begin{deluxetable}{cccccccc}
\tablecaption{Fit parameters to mean colors for face-on galaxies.}
\tablehead{
\colhead{mean color} & \colhead{$v_0$} & \colhead{$v_K$} & \colhead{$v_n$} &
\colhead{sigma} & \colhead{$s_0$} & \colhead{$s_K$} & \colhead{$s_n$}}
\startdata
$\nu_{(u-K)}$    &  1.67 & -0.27 &  0.26 &
$\sigma_{(u-K)}$ &  0.18 & 0.00 & 0.28\\
$\nu_{(g-K)}$    &  0.71 & -0.23 &  0.10  &
$\sigma_{(g-K)}$ &  0.25 &  0.07 & 0.08\\
$\nu_{(r-K)}$    &  0.30 & -0.20 &  0.02 &
$\sigma_{(r-K)}$ &  0.28 &  0.06 & 0.04\\
$\nu_{(i-K)}$    & 0.06 & -0.18 &  0.01 &
$\sigma_{(i-K)}$ &  0.28 &  0.06 & 0.02\\
$\nu_{(z-K)}$    & -0.02 & -0.17 &  -0.05 &
$\sigma_{(z-K)}$ &  0.28 &  0.04 & 0.01\\
$\nu_{(J-K)}$    & -0.01 & -0.01 &  0.01 &
$\sigma_{(J-K)}$ &  0.12 &  0.04 & 0.01\\
$\nu_{(H-K)}$    & -0.16 & -0.02 &  0.00 &
$\sigma_{(H-K)}$ &  0.13 &  0.06 &  0.01\\
\enddata
\tablecomments{This table lists the values of the parameters used to fit the mean 
distribution of face-on galaxy colors according to equation \ref{eq:color} and the 
standard deviation about that mean according to equation \ref{eq:sig}.}
\label{tab:color}
\end{deluxetable}

Another property that shows a strong difference between disk and elliptical galaxies 
is axis ratio.  Elliptical galaxies almost never have small axis ratios and galaxies 
with $b/a \le 0.55$ are $90\%$ disk galaxies in our sample.  This is understandable 
as disks are intrinsically thin and thus can have very small axis ratios when seen 
in projection, but ellipticals are close to spheroids whose intrinsic axis ratio $q_z$,
 is rarely less than 0.5 and will be larger than this when seen in projection.  
Thus axis ratio can be used to determine if a red concentrated galaxy is truly an 
elliptical galaxy or not.  Since red concentrated galaxies are almost always assumed 
to be early-type in the literature one may be skeptical of our claim that many of 
these galaxies are disk galaxies.  Thus we show an example of eighteen such galaxies 
in Figure \ref{fig:reddisks}.  These galaxies are selected to have $\nserc >  4.0$ 
and  $g-r > 0.75$ and $b/a < 0.5$.  These are the eighteen largest galaxies (in order to
make nice postage stamps) that meet the above criteria.  Clearly, highly-inclined, red, 
concentrated galaxies are not elliptical galaxies.  This is something that should be 
taken into account when one is trying to identify elliptical galaxies.  

If we select all galaxies with $\nserc  \le 3.0$ or $b/a \le 0.55$ this gives a sample that is 
$94\%$ disk galaxies and includes $70\%$ of all disk galaxies.  The disk galaxies missing from 
our sample are concentrated face-on disks. These galaxies will have the smallest inclination 
corrections and therefore are of the least concern for our application.  However, we need 
these galaxies to determine if the properties of concentrated disk galaxies are changing 
with inclination.  We discuss how we deal with this issue in \S\ref{sec:results}.

\subsection{Inclination from axis ratio}
\label{ssec:inc}
\begin{figure}[t]
   \centering
   \includegraphics[width=0.48\textwidth]{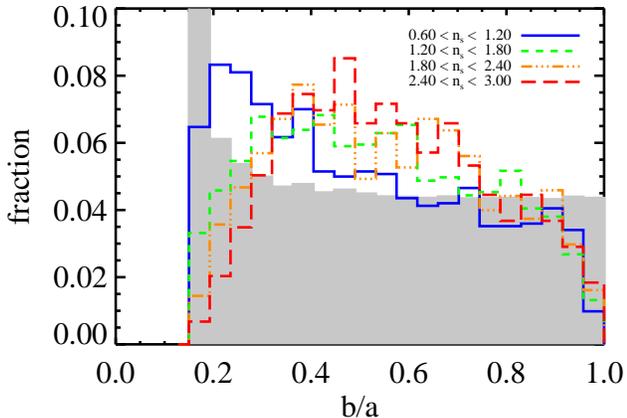} 
   \caption{The distribution of axis ratios, for galaxies in our sample in bins of Sersic 
   index.  The shaded region shows the theoretical expectation for the axis ratio of a 
   randomly inclined flattened spheroid with intrinsic axis ratio, $q_z = 0.15$.  For all 
   bins there is a deficiency of nearly circular galaxies compared to theory as is expected
   since any asymmetry in the light profile will move the isophotes away from 
   circular.  Galaxies with $\nserc > 1.2$ have nearly the same distribution of axis 
   ratios.  Only the least concentrated galaxies have an axis ratio distribution 
   similar to that of randomly oriented disks with finite thickness.  Even a small 
   bulge reduces the ellipticity of a galaxy's isophotes.  Nonetheless, it is evident 
   that axis ratio is mostly a measure of galaxy inclination.  
   }\label{fig:incdist}
\end{figure}

As mentioned above, the derivative in equation \ref{eq:main} is with respect to inclination, but 
inclination is not a directly measurable quantity.  Previous studies have dealt with this by assuming
that all disks have some average thickness and then turning the measured axis ratio into an inclination 
by
\beq
b/a = \sqrt{q_z^2+(1-q_z^2)\cos(\theta)},  
\eeq
where $q_z$ is the ratio between vertical and radial scale heights. This equation should 
hold if the three-dimensional light distribution is well fit by concentric ellipsoids and 
the disk is optically thin.  However, if one examines Figure \ref{fig:incdist}, which 
shows the distribution of observed axis ratios for galaxies with $\nserc \le 3.0$, binned 
by Sersic index, one sees that only the lowest $\nserc$ galaxies come close to having the 
distribution expected for randomly inclined disks with $q_z = 0.15$.  All bins show a 
deficit of face-on galaxies, which is easily understood as all sources of asymmetry in the 
galaxy will push one away from perfectly circular isophotes.  However, for galaxies with 
$\nserc > 1.2$ the existence of a bulge prevents these galaxies from having very low axis ratios.  
We do see that galaxies with $\nserc >  1.2$ show essentially the same distribution of
$b/a$, implying that the measured $b/a$ is mostly correlated with galaxy inclination.  
Because of these measurement issues we do not believe there is a reliable way to infer a 
galaxy's inclination from its axis ratio.  It is possible that bulge-disk decomposition 
fits may yield better results for the disk inclination and thus reduce this source of 
uncertainty.  For this paper we will simply work directly with axis ratio and reform 
equation \ref{eq:main} to be that a galaxy's intrinsic properties shouldn't depend on 
observed axis ratio.  

\section{Dependence of observed properties on axis ratio}
\label{sec:results}
We now turn to studying the dependence of observed galaxy properties on axis ratio.  We will start with
galaxy color, which is the property that we have found to have the strongest dependence on axis ratio.  
One complication when applying the method described in \S\ref{sec:method} to the property of 
color is that since color is a difference between two galaxy properties the correction found will not 
include any correction needed to the longer wavelength magnitude.  As the $K$-band is the longest 
wavelength we have access to we will consider $\lambda - K$ colors to minimize this effect.
Thus what is recovered in our case will be 
\beq
A_{\lambda} = A_{\lambda,tot} - A_K - A_{\lambda,\theta=0}
\eeq
where $A_{\lambda,tot}$ is the total attenuation in a given waveband, $A_K$ is any attenuation in the 
K-band and $A_{\lambda,\theta=0}$ is any attenuation in the face-on configuration.  This is why it is 
important to have infrared magnitudes for our sample, as we expect $A_K$ to be relatively small 
\citep{bd:01}.   In this section we will only show figures for $g-K$ colors, but our analysis 
is done for all six $\lambda -K$ colors we can produce.  To see the analogous figures for other 
wavebands please look at the supplemental online material or visit the website \url{www.galaxystats.com/intrinsic/}.

\begin{figure} [t]
   \centering
   \includegraphics[width=0.5\textwidth]{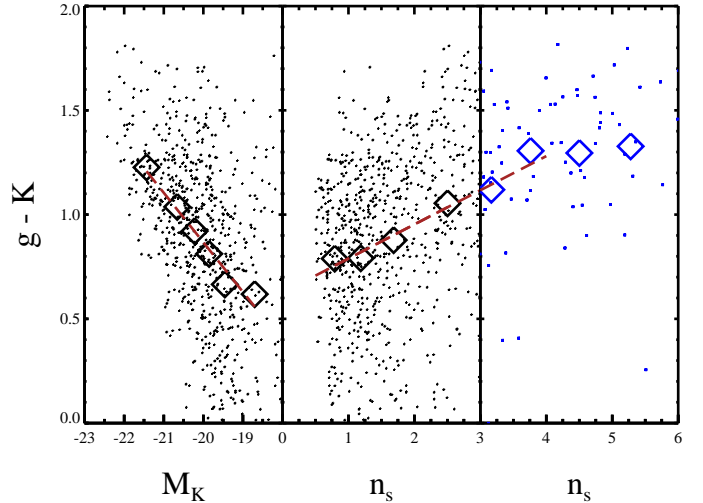} 
   \vspace{0.2cm}
   \caption{The g-K color of face-on galaxies ($b/a > 0.85$) is shown versus $M_K$ in the         left panel and $\nserc$ in the right two panels.  Diamonds show the mean value in a bin.  For $\nserc > 3.0$ (the right panel) we only plot galaxies visually classfied as disks to avoid any confusion with ellipticals.  The sample in the right panel is thus different then in the first two panels.  In spite of this, it is clear that galaxy color  is a function of luminosity and Sersic index.  In both cases the mean relationship is close to linear, with evidence for a flattening for $\nserc > 4.0$.
   }\label{fig:colorKn}
\end{figure}

In order to determine how a given galaxy's color depends on its inclination 
we need to know what that galaxy's color should be if we could see it face-on.  
Figure \ref{fig:colorKn} shows
the distribution of $g-K$ colors of face-on disk galaxies ($b/a > 0.85$) as a function of 
$K$-band luminosity and Sersic index.  The first two panels show all face-on galaxies with 
$\nserc \le 3.0$ while the third panel shows the face-on galaxies with $\nserc > 3.0$ that have 
been visually classified as disk galaxies by one of us (MRB).  We only include the visually 
classified
galaxies because we do not want to include elliptical galaxies, even though this means we are 
sampling a different luminosity range then the first two panels. We see that color depends on 
both Sersic index and luminosity which are not independent of one another.  The relationship 
looks close to linear for absolute magnitude and linear in Sersic index to a value of 
$\nserc = 4.0$ at which point it appears to flatten out.  Based on this we will assume that 
the dependence on $M_K$ and $\nserc$ are seperable and fit the mean galaxy color by a function 
of the form
\beq
\label{eq:color}
\nu_{(\lambda - K)} = v_0 + v_K (M_K + 20) + v_n \neff
\eeq
where $\nu_{(\lambda - K)}$ is the average $M_{\lambda} - M_K$ color of a face-on galaxy 
with absolute magnitude $M_K$ and Sersic index $\nserc$ and $\neff=\nserc$ for 
$\nserc \le 4.0$ and $\neff=4.0$ for $\nserc > 4.0$.  The values of $v_0, v_K$ and $v_n$ 
that best fit the data are shown in Table \ref{tab:color}.  One sees that the dependence of 
color on $\nserc$ is relatively weak except for $u-K$ colors, so this truncation for 
$\nserc \le 4.0$ generally has very little effect.

As can be seen in Figure \ref{fig:colorKn}, the dispersion about this mean also varies with 
$M_K$ and $\nserc$, so we also fit the standard deviation with the formula
\beq
\label{eq:sig}
\sigma_{(\lambda - K)} = s_0 + s_K(M_K + 20)  + s_n \neff.
\eeq
Again we use a value of $\neff=4.0$ for $\nserc > 4.0$ galaxies just to be consistent
with equation \ref{eq:color}.  Values of the fit parameters for dispersion are also given in 
Table \ref{tab:color}. One can see that the dependence on $\nserc$ is weak and whether 
or not one flattens the relationship at $\nserc = 4.0$ will have very little effect 
on our results.  In all cases, we find that these simple linear functions provides an 
adequate fit to the mean galaxy color and the standard deviation about this mean.

\begin{figure}[t] 
\begin{center}
\includegraphics[width=0.5\textwidth]{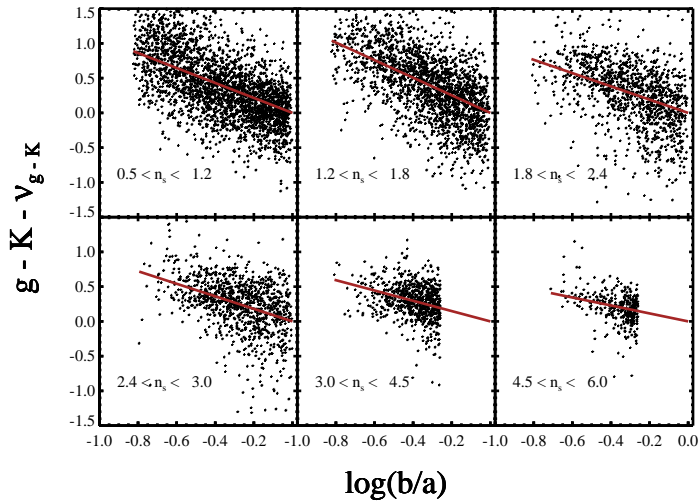}
\vspace{0cm}
\caption{The difference between a galaxy's $g-K$ color and the mean value of the $g-K$ color 
for a face-on galaxy with the same $M_K$ and $\nserc$ is shown as a function of axis ratio for six
different bins of $\nserc$.  We see in all cases, even for the highest $n$ bin, there is a clear 
trend for more inclined galaxies to be redder.  The relationship seems to be linear in 
$\log{b/a}$ with a slope that becomes shallower with higher $n$.  
}\label{fig:inc}
\end{center}
\end{figure}

With a model for the mean color of a face-on galaxy, we can now investigate how galaxy colors 
deviate from their mean face-on value as a function of axis ratio.  Figure \ref{fig:inc}
shows $g-K$ color minus $\nu_{(g-K)}$ versus $\log{b/a}$, for different bins of Sersic 
index. In all cases, a strong relationship is seen and one that seems to be roughly linear 
in $\log{b/a}$.  The large intrinsic scatter in galaxy color makes it difficult to judge whether 
a more complicated dependence on $\log{b/a}$ is warranted.  As a linear relationship seems 
adequate, we will express the attenuation in the form

\beq
\label{eq:a}
A_{\lambda} = -\gamlam  \log{b/a}.
\eeq
Note that this is different than what has been done historically where the attenuation 
is considered to be linear in $\log{(\cos\theta)}$ (equation \ref{eq:old}). Here we are 
expressing the attenuation in terms of the observed axis ratio. It would be very 
interesting to compare this to the dependence of attenuation on observed axis ratio in 
radiative transfer simulations.

\begin{figure}[t] 
   \centering
   \includegraphics[width=0.5\textwidth]{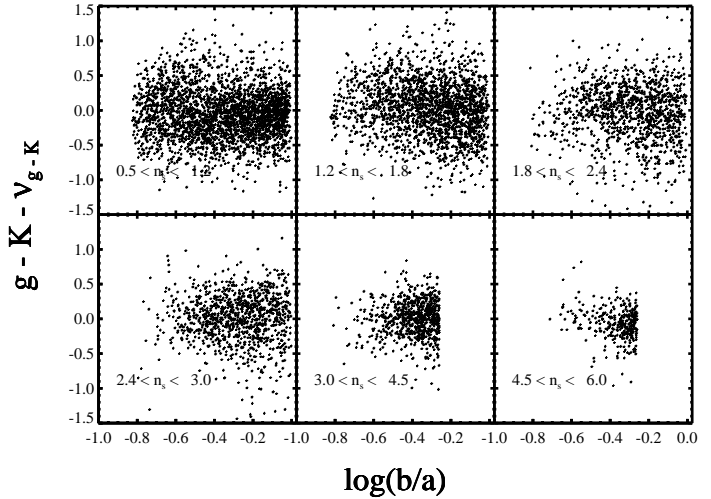} 
   \vspace{0cm}
   \caption{The difference between a galaxy's intrinsic $g-K$ color (corrected for 
   inclination) and the mean value of the $g-K$ color for a face-on galaxy with the 
   same $M_K$ and $\nserc$ is shown as a function of axis ratio for six different 
   bins of $\nserc$.  We see that applying the inclination correction removes
   the dependence of color on inclination seen in Fig \ref{fig:inc}.
  }\label{fig:inccor}
\end{figure}

Examining Figure \ref{fig:inc}, we see that $\gamlam$ (the slope of the line) varies with 
Sersic index.  Creating plots like Figure \ref{fig:inc} and dividing the sample by other galaxy 
properties, we find that the correlation between color and axis ratio is strongest when binned
by $\nserc$ or $M_K$.  Therefore we uncover that $\gamlam$ depends on at least these two parameters.
We start with the simplest assumption, that $\gamlam$ can be determined from a 
linear combination of these parameters:
\beq
\label{eq:gamma}
\gamlam = \alpha_0+ \alpha_K (M_K  + 20) + \alpha_n \neff,
\eeq
where the parameters $\alpha_0, \alpha_n$ and $\alpha_z$ depend on wavelength. Again we will 
assume that dependence flattens off for $\nserc > 4.0$.  Note that 
previous studies have only considered a linear dependence on luminosity, not Sersic-index.
Now that we have a model for the attenuation we can investigate the best fit parameters 
for equation \ref{eq:gamma}.  To do this we use a Monte Carlo Markov Chain (MCMC) to minimize 
the statistic,    
\beq
\label{eq:chi}
\chi^2 = \sum{\left[{{(M_{\lambda}-A_{\lambda} - M_{K}) - \nu_{(\lambda - K)} }\over
{\sigma_{(\lambda - K)}}}\right]^2}
\eeq
where $A_{\lambda}$ is given by equations \ref{eq:a} and \ref{eq:gamma}.  The resulting best 
fit parameters are shown in Table \ref{tab:params}. Figure \ref{fig:inccor} shows the 
effect of using these corrections to determine the intrinsic colors of galaxies in our
sample.  Plotted is the intrinsic g-K color versus axis ratio for our disk galaxy sample
We see that, unlike in Figure \ref{fig:inc}, there is not a discernible correlation between
galaxy color and axis ratio.  Thus, it seems that our simple linear functions are sufficient
to describe the behavior of the attenuation.  When Figure \ref{fig:inccor} is made for the 
other 5 wavebands, a similar lack of correlation is seen.    

Table \ref{tab:params} also gives the mean value of $\gamlam$, the mean value of 
the attenuation and the maximum attenuation of any galaxy in our sample for each 
waveband.  We see that in the $g$ band the attenuation can be as large as 1.25 
magnitudes, but that the mean attenuation is $0.28$ magnitudes.  These values 
decrease substantially with longer wavebands reaching a mean attenuation of 
only a few hundredths of a magnitude for the near infrared bands.  We expect 
any attenuation in the $K$-band to be less than this and thus negligible.  

\begin{figure}[t] 
   \centering
   \includegraphics[width=0.5\textwidth]{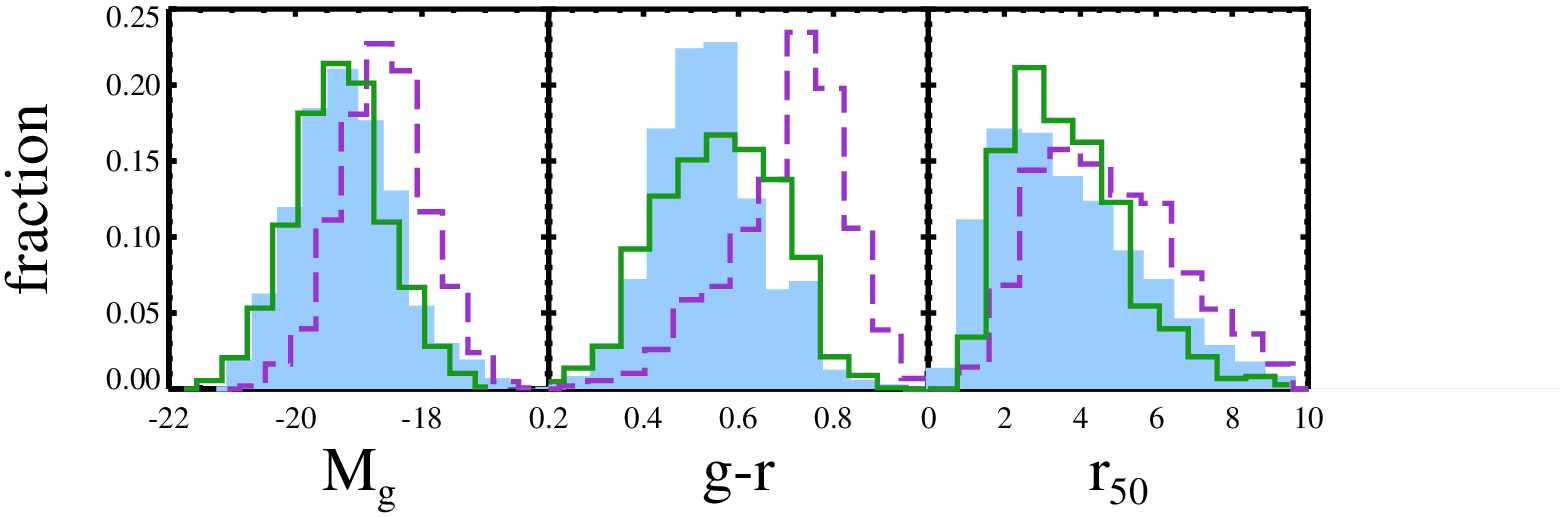}
   \vspace{-1cm}
   \caption{The left most panel shows the distribution of observed 
   $g$-band magnitudes for face-on (shaded) and edge-on (dashed line) galaxies.
  Also shown is the distribution of intrinsic (corrected) 
magnitudes for the edge-on (solid line) galaxies. One sees that the 
distributions of observed magnitudes is very different, but that 
once an inclination correction has been applied the distributions become 
almost identical.  The middle panel shows distributions of $g-r$ color, with the 
same line-styles as the left most panel.  Again the inclination corrections bring 
the edge-on galaxies into agreement with the face-on ones. However, the shape
of the corrected color distribution differs from the face-on one, suggesting 
that their may be dependence on another parameter needed for the correction. 
The right most panel shows the distribution of observed galaxy half light radii, 
$r_{50}$, again with the same line styles indicating observed face-on, observed 
edge-on and intrinsic edge-on galaxies. In all three panels only galaxies with 
$\nserc \le 3.0$ are used to insure a fair comparison.}
   \label{fig:res}
\end{figure}

As a semi-independent check of our results, we compare the distribution of $M_g$
magnitudes and $g-r$ colors for $\nserc \le 3.0$ galaxies in our sample.  We
restrict ourselves to these low concentration galaxies so that we can compare 
face-on ($b/a \ge 0.85$) and edge-on ($b/a \le 0.30$) disk galaxies without having 
contamination from ellipticals.  The distributions of these quantities for 
for face-on (shaded), uncorrected edge-on (dashed line) and corrected edge-on 
(solid line) galaxies is shown in Figure \ref{fig:res}.  One sees that the 
distributions of face-on and edge-on galaxies is significantly different.   
The mean $M_g$ magnitude and $g-r$ color of face-on galaxies are $-19.2$ and 
$0.54$, respectively; for edge-on galaxies the values are $-18.6$ and $0.71$.      
This significant difference is dramatically reduced when considering the 
corrected magnitudes of the edge-on galaxies, which have mean values of $-19.3$
and $0.56$, respectively.  Examination of the histograms in Figure \ref{fig:res}
shows that while corrected $g-K$ colors have nearly the 
same mean as the face-on colors, there is some difference in the shape of the
distribution.  This seems to be an indication that an additional parameter 
or possibly a cross term is needed to fully describe the correction, though 
this is difficult to tell since there is a small bias introduced because edge-on 
galaxies are included in the sample based on their observed and not their 
intrinsic magnitudes.  Either way, these differences are relatively small and 
the overall success of our inclination corrections is evident.  

Having determined the correction for magnitudes we now turn to another galaxy 
property that shows inclination dependence, galaxy size.  Our measured half 
light radii $r_{50}$ are only in the r-band, though it would be very informative
for dust modeling to look at the radial dependence of attenuation in different
wavebands.  Examining the face-on sizes of galaxies in our sample we see that 
they depend on K-band luminosity \citep[i.e., the Kormandy relation,][]{korm:77}, 
but show no dependence on sersic index.  We find the mean face-on half light radii are 
well fit by
\beq
\label{eq:size}
\log r_{50} = 0.5 - 0.13(M_K +20).
\eeq
The distribution around this value is roughly lognormal as seen in previous studies
\citep{dl:00,shen:03} so we will perform our correction in $\log{r_{50}}$.   Comparing
edge-on to face-on galaxies we find that galaxies with $\nserc \le 2.0$ have larger half 
light radii when they are inclined.  For galaxies with $\nserc > 2.0$ we see no effect, probably because of the prominence of bulges in these galaxies.  This increase of the
sizes of inclined galaxies is what one would expect if there is more
dust in the inner parts of galaxies causing greater attenuation in the center. However,
as noted earlier, this may also be an artificate of the data pipeline used to determine 
galaxy size or a combination of both dust and the data pipeline.
We parametrize the effect for galaxies by the formula
\beq
\label{eq:asize}
\log r^i_{50} = \log r^o_{50} +\beta_r(\log b/a )
\eeq
where $r^i_{50}$ and $r^o_{50}$ are the intrinsic and observed half light radii, 
respectively, and $\beta_r$ is the strength of the effect in the $r$-band (analogous 
to $\gamlam$).  Minimizing $\chi^2$ gives values of $\beta_r =0.2$.   
We find that we do not need luminosity or sersic index dependence to bring the edge-on 
galaxies into agreement with the face-on galaxy size distribution.  This can be seen in 
the third panel of Figure \ref{fig:res} which shows the face-on and the corrected edge-on galaxy size distributions to be in good agreement.  It will be interesting to continue this analysis for the full SDSS galaxy catalogue as we plan to in a following paper.   Having  determined corrections for observed magnitudes and sizes of galaxies in our sample,  we now turn to comparing our results to previous determinations of the inclination correction in the literature.

\begin{deluxetable}{ccccccc}
\tablecaption{Inclination correction parameters for different wave bands}
\tablehead{
\colhead{Band} & \colhead{$\alpha_0$} & \colhead{$\alpha_K$} & \colhead{$\alpha_n$} & 
\colhead{max $\gamlam$} & \colhead{mean $A_{\lambda}$} & \colhead{max $A_{\lambda}$}}
\startdata
$A_u$ &  1.79 & -0.43 & -0.28 &  1.31 &  0.39 &  1.88\\
$A_g$ &  1.38 & -0.19 & -0.22 &  0.94 &  0.28 &  1.24\\
$A_r$ &  1.02 & -0.09 & -0.12 &  0.78 &  0.23 &  0.88\\
$A_i$ &  0.90 & -0.05 & -0.14 &  0.60 &  0.18 &  0.70\\
$A_z$ &  0.56 &  0.01 & -0.01 &  0.53 &  0.15 &  0.46\\
$A_J$ &  0.31 & -0.02 & -0.07 &  0.17 &  0.05 &  0.24\\
$A_H$ &  0.15 &  0.01 & -0.05 &  0.05 &  0.02 &  0.11\\
\enddata
\tablecomments{The table shows the parameters used to fit for $\gamlam$ to 
determine the attenuation in different bands according to eqns. 
\ref{eq:a} and \ref{eq:gamma}.  Also shown are the mean and maximum attenuation 
in each waveband and the maximum value of $\gamlam$.}
\label{tab:params}
\end{deluxetable}

\section{Comparison to Previous Work}
\label{sec:compare}

In comparing our results to previous work it is important to take into account three
differences in our treatment.  One, our value of $\gamlam$ is multiplied by $\log{b/a}$
instead of $\log{(\cos{\theta})}$.  Two, we explore $\gamlam$ dependence on both $M_K$ and $n_s$, while earlier studies only considered dependence on luminosity.  Finally, 
we include all types of galaxies in our sample instead of focusing on just late 
type galaxies.

\begin{figure}[t] 
   \centering
   \includegraphics[width=0.5\textwidth]{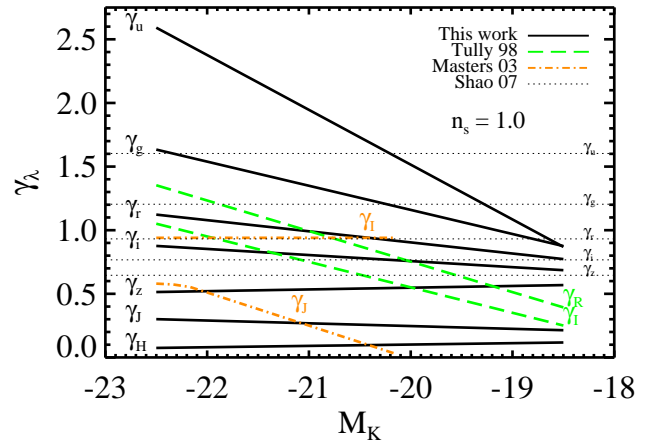}
   \vspace{0cm}
   \caption{The value of $\gamlam$ for spiral galaxies ($\nserc =1$) as a function 
   of $M_K$ is shown in comparison to other results from \citet[dashed lines]{tully:98},
   \citet[dash-dot lines]{mgh:03} and \citet[dotted lines]{shao:07}. Note that none of 
   the wavebands are the same (except for the $J$-band which are all from 2MASS)
   and that the methods used to determine $\gamlam$ all differ.}
   \label{fig:gamcomp}
\end{figure}

Our values of $\gamlam$ for an $\nserc=1.0$ galaxy are shown as a function of galaxy 
luminosity in Figure \ref{fig:gamcomp}.  One sees that aside from in the $u$-band, 
the dependence on luminosity is fairly weak.  For comparison, we show the values 
determined by \citet{tully:98,mgh:03} and \citet{shao:07}.  Note that 
\citeauthor{shao:07} found no evidence for luminosity dependence, so thier fit 
assumes this and that $\gamlam$ has a power law dependence on wavelength.  Also,
to compare to \citeauthor{tully:98}, we have converted the measured dependence to 
$M_K$ by subtracting the mean $r-K$ and $i-K$ colors of our galaxy sample. We see
that the determinations of $\gamlam$ from all studies have similar values, that is
they all agree with one another at some luminosity. However, the dependence on 
luminosity shows a wide range of behaviors. There are many possible reasons for 
these differences.  One is definitely the assumptions that have been made, 
for example the $\gamma_{I}$ value from \citeauthor{mgh:03} assumes no 
luminosity dependence, while the $\gamma_J$ values is fit to a broken line.  
Clearly, we expect that we will get different results because we have fit 
to both luminosity and Sersic index. Other issues that may be important are 
the sample selection used in the study, the conversion from b/a to inclination
done in the other studies, and which property is being correlated with 
inclination.  \citeauthor{tully:98} and we both used galaxy color, 
while \citeauthor{mgh:03} looked at magnitude in an isophotal radius and 
\citeauthor{shao:07} focused on the galaxy luminosity function.  In particular, 
as we have shown in \S\ref{sec:results}, not only is the total luminosity of a 
galaxy dependent on axis ratio, but also the half light radius is too.  
Since isophotal radius depends on both it may not be surprising that we arrive at 
different results than \citeauthor{mgh:03}  These intricacies will require 
further study, which would be most enlightening when combined with radiative transfer 
simulations. 

\begin{figure}[t] 
   \centering
   \includegraphics[width=0.5\textwidth]{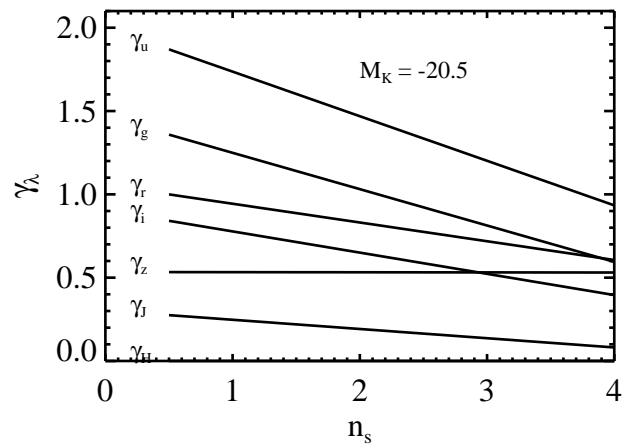} 
   \vspace{0cm}
   \caption{The value of $\gamlam$ as a function of Sersic index. $\gamlam$ shows
   a strong dependence on Sersic index in all bands except for the $z$-band.  
   $\gamma_z$ shows almost no dependence on $\nserc$, which becomes unphysical 
   when $\gamma_z \ge \gamma_i$.  This seems to imply that our assumption that 
   $\gamlam$ is linear in $\nserc$ should be modified. Unfortunately, unlike in 
   Fig. \ref{fig:gamcomp} we can not compare our results as other studies have
   not calculated $\gamlam$ as a function of Sersic index.
   }\label{fig:gamserc}
\end{figure}

The dependence of $\gamlam$ on Sersic index is shown in Figure \ref{fig:gamserc}. 
In general, we find that the dependence on Sersic index is stronger than on 
$K$-band magnitude. We have fit each band independently so the unphysical situation 
that arises in Figure \ref{fig:gamserc} can occur where $\gamma_z$ becomes 
larger than $\gamma_i$ at high Sersic index.  This is of course undesirable, 
but our approach is to investigate what we can learn from the data instead 
of coming in with theoretical prejudices.  One can easily find an acceptable 
solution to this by requiring that $\gamlam$ is always smaller in longer 
wavelengths.  That would mean that in Figure \ref{fig:gamserc}, $\gamma_z$ 
would take on the value of $\gamma_i$ where now $\gamma_z > \gamma_i$.  One way 
to improve in this area would be to fit multiple wavelengths at the same time, 
under some assumptions about how the attenuation at different wavelengths is 
related.

\begin{figure}[t] 
   \centering
   \includegraphics[width=0.5\textwidth]{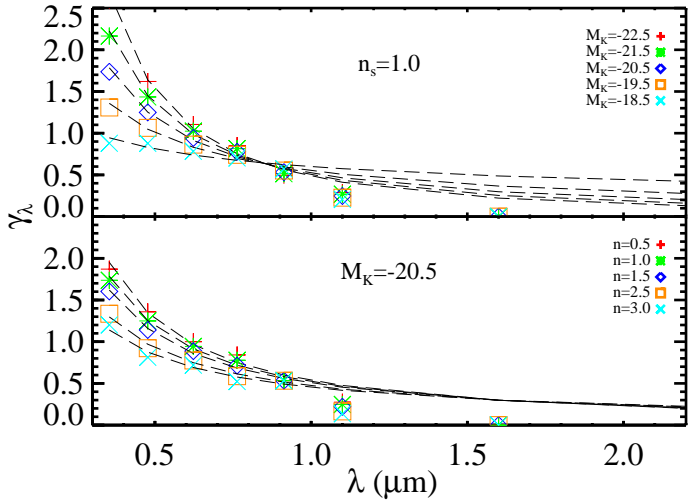} 
   \vspace{0cm}
   \caption{The value of $\gamlam$ as a function of wavelength for various values of $M_K$ 
   at fixed $\nserc$ (top panel) and for various values of $\nserc$ at fixed $M_K$ (bottom 
   panel). The lines are power law fits to the 5 values for the SDSS bands.  One sees that
   though the first 5 bands seem to be well fit by a power law the measured values of 
   $\gamma_J$ and $\gamma_H$ are always well below the extrapolation of this curve.   
   }\label{fig:gamlam}
\end{figure}

Since we fit each waveband independently and since $\gamlam$ depends
on both $M_K$ and $\nserc$, the wavelength dependence on $\gamlam$, and thus 
the attenuation, varies with $M_K$ and $\nserc$. Figure \ref{fig:gamlam} shows 
the wavelength dependence of $\gamlam$ for fixed values of $M_K$ and $\nserc$.
We see that generally a power law is a good fit to $\gamlam$ in the SDSS 
wavebands.  The index of the power law seems to increase with $M_K$ for fixed 
$\nserc$, while at fixed $\nserc$ the slope of the index of the power law remains
almost unchanged and only the amplitude of $\gamlam$ increases with decreasing 
$\nserc$.  The power law index ranges from $\sim 0.5$ to $\sim 1.5$. These 
relationships should be very useful in understanding the 
distribution of dust in these galaxies. We see in all cases $\gamlam$ in
the near infrared bands (J and H) falls below the power law fit.  It is clear 
from this that the amount of attenuation in the K-band, not included in our 
calculations, is negligible.  

We conclude from this that the corrections we find are in general agreement with
previous results and that differences are probably primarily due to our including 
the dependence on Sersic index.  We now turn to a preliminary exploration of how 
galaxy properties differ when considering the intrinsic instead of 
observed values.    

\section{Intrinsic galaxy properties compared to observed ones}
\label{sec:changes}

In this section we turn to comparing intrinsic galaxy properties to observed 
ones.  A complete discussion of this subject is beyond the scope of this work. And as we 
have mentioned earlier, we will not be discussing the space density of objects because we 
do not have a flux limited sample.  Instead we will focus on how galaxy photometric quantities
change in our sample when going from observed to intrinsic properties.
The overall effect is fairly modest, as can be seen from Table \ref{tab:params}, the 
average attenuation in the $g$-band is $A_g=0.28$. This will have small effects on 
the galaxy luminosity function and correlation function, essentially a small shift in
their amplitudes. However, the inclination correction is not uniform, 
and while many galaxies have no correction others are brightened by up to 1.25 magnitudes 
in the $g$-band.  Thus the most notable effect will be in comparing the properties of spiral
and elliptical galaxies.  We will focus on the differences between observed and
intrinsic galaxy color as seen in the color magnitude diagram, the determination
of stellar masses, and the determination of photometric redshifts, all measurements 
that are based on galaxy color.  First though, we start with discussing survey completeness.

\subsection{survey completeness}
When information is presented from a galaxy survey it is usually taken from a flux 
limited catalog.  The SDSS spectroscopic sample is complete to an apparent 
magnitude of $m_r = 17.7$.  This can be converted to an equivalent completeness in 
observed absolute magnitude at some distance.  But this does not translate into an 
equivalent completeness in intrinsic magnitudes.  Thus, when one looks at a flux limited 
catalog, there are galaxies that are intrinsically identical to galaxies in the catalog, 
but are excluded because of the inclination they happen to be viewed at.  When 
discussing the distribution of galaxy properties or connecting galaxies to dark matter 
halos this is not desirable.  In intrinsic magnitudes, the survey is only complete to 
the maximum attenuation brighter than the observed magnitude.  So in the $r$-band, 
where the maximum attenuation is $\sim0.9$ magnitudes, this corresponds to 
$m_r =16.8$ a much shallower survey.  

If one wants an intrinsically complete survey it is not necessary to obtain spectra
for every galaxy fainter than the survey limit down to this maximum attenuation.  Instead
one only needs to target galaxies where 
\beq
m_r < 17.7 + {\rm max}(\gamma_r)*\log{b/a} 
\eeq
a much smaller sample of galaxies. 
If one determines Sersic indeces before targeting the galaxies then ${\rm max}(\gamma_r)$
would equal the maximum value of $\gamma_r$ for that value of $\nserc$. Without the
Sersic indeces (or some equivalent measure) one would have to use the maximum value of 
$\gamma_r$ which for our sample is 1.1.  To give an example, we use the DR4 sample of 
SDSS, which does not have determinations of Sersic indeces, but has measured axis 
ratios, though different then the ones we have been using in this paper 
(see \S\ref{sec:data}).  These axis ratios are measured with exponential fits 
(labeled ab\_exp in the catalog), but are reasonably close to the ones we have been
using which are based on elliptical Sersic fits.  In this sample there are 92,752 
galaxies with $m_r < 16.8$.  Going $0.9$ mag deeper to $m_r < 17.7$ increases the 
sample by 242,103 galaxies. However, if the axis ratio of the galaxy is considered 
then there are only 89,855 galaxies that may be brighter than 16.8 mags after correcting
for inclination.  Thus taking the axis ratio into account reduces the required follow up 
spectra by $63\%$.  Future surveys may want to consider axis-ratio along with apparent magnitude when determine their galaxy sample. In this way, samples limited to some intrinsic magnitude can be compiled which will greatly facilitate comparisons to theory and other observations.

\subsection{color-magnitude relationship}
\begin{figure}[t] 
   \centering
   \includegraphics[width=0.5\textwidth]{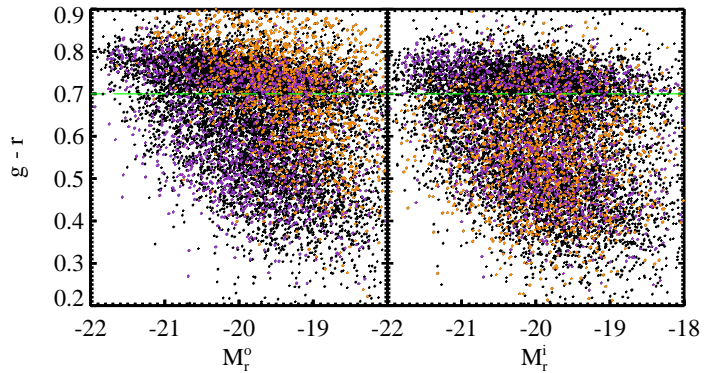} 
   \vspace{0cm}
   \caption{The observed color-magnitude diagram (left panel) and the intrinsic 
   color-magnitude diagram (right panel).  In the observed color magnitude diagram
   there are many galaxies redder than the red sequence.  These galaxies are mostly
   removed in the intrinsic color-magnitude diagram.  Face-on ($b/a > 0.85$) and 
   edge-on ($b/a < 0.35$) galaxies are color coded by purple and orange, respectively.
   In the left panel they occupy a completely different region of the diagram, but in the 
   right panel they have similar distributions.     
   }\label{fig:cm}
\end{figure}

One area where we see a significant difference between observed and intrinsic properties 
is in the color-magnitude diagram.  Figure \ref{fig:cm} shows a color-magnitude diagram  
for observed and intrinsic magnitudes for our sample.  There is no $1/V_{max}$ weighting 
as explained earlier, which would increase the difference between the two plots. Clearly,
the distribution of points has changed in the two panels. In the observed color-magnitude 
diagram $46\%$ of galaxies have $g-r \ge 0.7$; for the intrinsic color-magnitude diagram 
this decreases to $32\%$.  Therefore, in observed colors the split between red and blue 
is roughly 1:1, in intrinsic color this changes to 1:2.  Assertions about how the galaxy 
population is evolving based on the number of red or blue galaxies \citep{faber:05,bell:07} 
should probably look at changes in intrinsic instead of observed properties if they want to 
be sure that reddening isn't a large part of the effect they are measuring.

\subsection{stellar masses}
\begin{figure*}[t] 
   \centering
   \includegraphics[width=\textwidth]{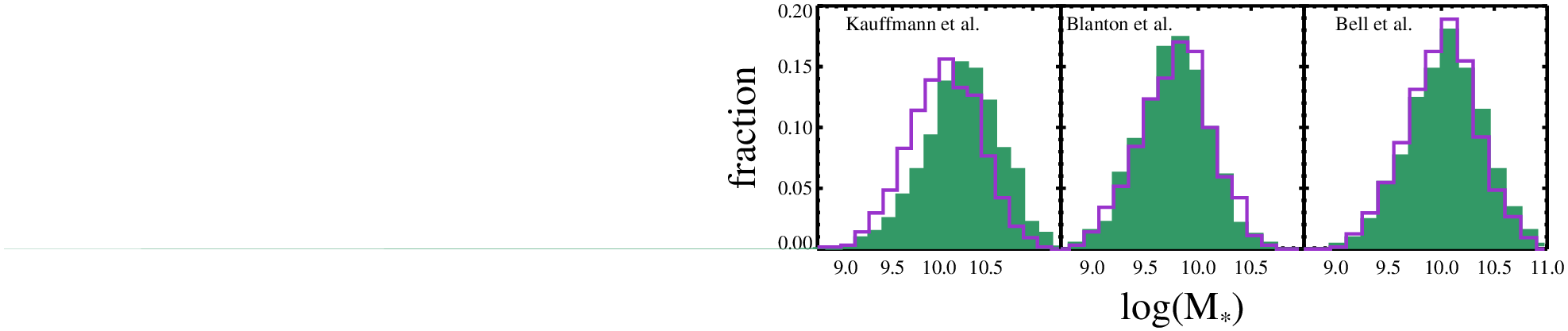} 
   \vspace{-2.5cm}
   \caption{The distribution of stellar masses in our sample for face-on (shaded)
   and edge-on (line) galaxies with $\nserc \le 3.0$.  The left panel shows
   stellar masses determined by \citet{kauf:03}, the middle panel by \citet{blan:03a}
   and the right panel by \citet{bell:03a}.  We see that in the left panel the edge-on
   galaxes are not corrected enough, in the middle panel they are corrected too much
   and for the right panel the correction is just right.  Note that this only tests
   that there is not inclination dependence in the determination, not that the
   stellar masses are correct.
   }\label{fig:sm}
\end{figure*}

One intrinsic galaxy property that we have not discussed yet is stellar mass.  Stellar
mass is not directly observable, but there are a number of methods that try to estimate 
the stellar mass based on spectroscopy or photometry, \citep{bell:03a,blan:03a,kauf:03}.
Clearly, stellar mass is an intrinsic property - the mass of a galaxy shouldn't change
under rotation.  To check that this is the case then, one would like to see that equation
\ref{eq:main} holds for stellar mass.  Figure \ref{fig:sm} shows histograms of 
the stellar mass for face-on ($b/a \ge 0.85$, shaded) and edge-on ($b/a \le 0.3$, line) 
galaxies using the mass estimates of \citet{kauf:03}, \citet{blan:03a} and \citet{bell:03a}.
We see that in the estimates of \citeauthor{kauf:03} edge-on galaxies are slightly less
massive then face-on galaxies, that for \citeauthor{blan:03a} edge-on galaxies are slightly
more massive then face-on galaxies and that for \citeauthor{bell:03a} there is remarkably
good agreement between the two.  We should note that the method of \citeauthor{kauf:03} 
is based on spectra which, for the low redshift sample we are considering will only 
probe the inner parts of a galaxy and therefore is more likely to be in error.  Also,
we are only testing that the determined stellar mass does not depend on 
inclination not that it is the correct value of the stellar mass.  The point of this
exercise is just to demonstrate that one should check that intrinsic quantities 
actually don't show dependence on inclination.    

\subsection{photometric redshifts}
Finally, we turn to the issue of determining photometric redshifts
\citep[e.g.,][]{hogg:98} for galaxies, or clusters found by identifying 
red-sequence galaxies \citep[e.g][]{koes:07}, which is a subset of photometric 
redshifts.  In recent years algorithms to determine a galaxy's redshift 
only using photometric redshifts have become very successful.  These methods rely on
the fact that there is a limited family of intrinsic SEDs that galaxies have so that 
with enough wavelength coverage it is possible to determine both the galaxy's SED and
redshift with only photometric data.  However, as we have demonstrated in this paper,
the observed SED of a galaxy changes with inclination.  It may still be possible to 
disentangle the intrinsic SED, reddening and redshift with observations in enough
wavebands, but in cases where an axis ratio can be measured, photometeric redshifts 
ignore useful information.  Thus we strongly recommend that in future 
determinations of photometric redshifts axis ratio be used to help reduce uncertainties.
 
\section{Conclusion}
\label{sec:conc}

This paper has focused on the intrinsic properties of galaxies, which are routinely
determined for Tully-Fisher studies \citep[e.g.,][]{piza:07}, but not for other 
statistical studies of galaxies.  Intrinsic properties are invariant under changes 
in viewing angle, unlike observed properties which would change if we 
could view a galaxy from a different vantage point.  Our main goal in this paper 
has been to clarify the difference between observed and intrinsic properties and 
to suggest how observed properties can be converted into intrinsic ones.

The method we use in this paper is to identify an observed galaxy property 
that shows a correlation with axis ratio and then apply the necessary correction to remove      
this correlation.  We find that both color and size show correlations with axis ratio. We are 
able to remove these correlations using simple linear formula that depend on $K$-band magnitude
and Sersic index.  We therefore can construct a galaxy catalog with intrinsic value for galaxy 
size and magnitude.  

There are many distributions and relationships that should be reconsidered in terms of the 
intrinsic properties instead of the observed ones.  We can not cover all of these in a paper
of this length but we highlight a few points to suggest how things may change.  We focus
on the color magnitude diagram as an example.  The observed color magnitude diagram shows
a number of galaxies redder than the red sequence and a lack of bright blue galaxies.  When 
we plot the intrinsic color magnitude diagram these ultra red galaxies are mostly removed and 
there are many more bright blue galaxies.  We find that the ratio of blue to red galaxies changes
from 1:1 to 2:1 for galaxies with absolute luminosities $-23.75 \ge M_K > -17.75$, 
a significant change.  

In a following paper, we apply the insights we have gained here to producing inclination
corrections for the full SDSS catalog.  Having intrinsic properties for this flux limited
catalog will then allow us to address how volume densities are effected when going from
observed to intrinsic quantities.

\acknowledgments

AHM is grateful for many fruitful conversations with Martin Weinberg during the conception of this project. We would like to thank Eric Bell, James Bullock and Dan McIntosh for helpful comments on earlier drafts of this paper.  AHM was partially supported by GRTI-CT05AGR9001 and PSC-CUNY-60021-36-37 grants from CUNY.   MRB was partially supported during this work by grants NASA-06-GALEX06-0030, NSF-AST-0607701, and Spitzer G03-AT-30842M.

Funding for the SDSS and SDSS-II has been provided by the Alfred P. Sloan Foundation, the Participating Institutions, the National Science Foundation, the U.S. Department of Energy, the National Aeronautics and Space Administration, the Japanese Monbukagakusho, the Max Planck Society, and the Higher Education Funding Council for England. The SDSS Web Site is http://www.sdss.org/.

The SDSSf is managed by the Astrophysical Research Consortium for the Participating Institutions. The Participating Institutions are the American Museum of Natural History, Astrophysical Institute Potsdam, University of Basel, University of Cambridge, Case Western Reserve University, University of Chicago, Drexel University, Fermilab, the Institute for Advanced Study, the Japan Participation Group, Johns Hopkins University, the Joint Institute for Nuclear Astrophysics, the Kavli Institute for Particle Astrophysics and Cosmology, the Korean Scientist Group, the Chinese Academy of Sciences (LAMOST), Los Alamos National Laboratory, the Max-Planck-Institute for Astronomy (MPIA), the Max-Planck-Institute for Astrophysics (MPA), New Mexico State University, Ohio State University, University of Pittsburgh, University of Portsmouth, Princeton University, the United States Naval Observatory, and the University of Washington. 

This publication makes use of data products from the Two Micron All Sky Survey, which is a joint project of the University of Massachusetts and the Infrared Processing and Analysis Center/California Institute of Technology, funded by the National Aeronautics and Space Administration and the National Science Foundation.

\bibliographystyle{apj}

\bibliography{everything}

\end{document}